\documentclass{naturewithfigs}

\usepackage{txfonts}
\usepackage{graphicx}
\usepackage{amssymb} 
\usepackage{hyperref}

\def\apj{ApJ}

\title{A wide and collimated radio jet in 3C\,84 on the scale of a few hundred gravitational radii}

\author{G.~Giovannini$^{1,2}$, T.~Savolainen$^{3,4,5}$, M.~Orienti$^{2}$, M.~Nakamura$^{6}$, H.~Nagai$^{7}$,  M.~Kino$^{15,8}$, M.~Giroletti$^{2}$, K.~Hada$^{8}$, G.~Bruni$^{5,2,9}$, Y.Y.~Kovalev$^{10,11,5}$, J.M.~Anderson$^{12}$, F.~D'Ammando$^{1,2}$, J.~Hodgson$^{13}$, M.~Honma$^{8}$, T.P.~Krichbaum$^{5}$, S.-S.~Lee$^{13,16}$, R.~Lico$^{1,2}$, M.M.~Lisakov$^{10}$, A.P.~Lobanov$^{5}$, L.~Petrov$^{14,11}$, 
B.W.~Sohn$^{13,16}$, K.V.~Sokolovsky$^{10,17,18}$, P.A.~Voitsik$^{10}$, J.A.~Zensus$^{5}$, S.~Tingay$^{19}$}

\begin{document}

\maketitle

\begin{affiliations}
 \item Dipartimento di Fisica e Astronomia, Universit\`a di Bologna, via Gobetti 93/2, 40129 Bologna, Italy
 \item INAF -- Istituto di Radio Astronomia, via P. Gobetti 101, I-40129 Bologna, Italy. email:ggiovann@ira.inaf.it
 \item Aalto University Department of Electronics and Nanoengineering, P.O. Box 15500, FI-00076 Aalto, Finland
 \item Aalto University Mets\"ahovi Radio Observatory, Mets\"ahovintie 114, FI-02540 Kylm\"al\"a, Finland
 \item Max-Planck-Institut f\"ur Radioastronomie, Auf dem H\"ugel 69, 53121 Bonn, Germany
  \item Institute of Astronomy and Astrophysics, Academia Sinica, 11F Astronomy-Mathematics Building, AS/NTU No. 1, Taipei 10617, Taiwan
  \item National Astronomical Observatory of Japan, Osawa, Mitaka, Tokyo 181-8588, Japan
 \item Mizusawa VLBI Observatory, National Astronomical Observatory of Japan, 2-21-1, Osawa, Mitaka, Tokyo 181-8588, Japan
 \item INAF -- Istituto di Astrofisica e Planetologia Spaziali, via Fosso del Cavaliere 100, 00133 Roma, Italy 
 \item  Astro Space Center of Lebedev Physical Institute, Profsoyuznaya 84/32, 117997 Moscow, Russia
 \item Moscow Institute of Physics and Technology, Dolgoprudny, Institutsky per., 9, Moscow region, 141700, Russia
 \item Helmholtz Centre Potsdam, GFZ German Research Centre for Geosciences, 14473 Potsdam, Germany
 \item Korea Astronomy and Space Science Institute (KASI), 776 Daedeokdae-ro, Yuseong-gu, Daejeon 305-348, Korea
 \item Astrogeo Center, 7312 Sportsman Dr., Falls Church, VA 22043, USA
 \item Kogakuin University, Academic Support Center, 2665-1 Nakano, Hachioji, Tokyo 192-0015, Japan
 \item Korea University of Science and Technology, 217 Gajeong-ro, Yuseong-gu, Daejeon 34113, Korea
\item IAASARS, National Observatory of Athens, Vas.~Pavlou \& I.~Metaxa, 15236 Penteli, Greece
 \item Sternberg Astronomical Institute, Moscow State University, Universitetskii~pr.~13, 119992 Moscow, Russia
 \item Curtin Institute of Radio Astronomy, Curtin University, Bentley WA6102, Australia
\end{affiliations}

\newpage

\begin{abstract}
Understanding the launching, acceleration, and collimation of jets powered by active galactic nuclei remains an outstanding problem in relativistic astrophysics\cite{Boettcher2012}. This is partly because observational tests of jet formation models suffer from the limited angular resolution of ground-based very long baseline interferometry that has thus far been able to probe the transverse jet structure in the acceleration and collimation zone of only two sources\cite{Asada2012,Boccardi2016}. Here we report radio interferometric observations of 3C\,84 (NGC\,1275), the central galaxy of the Perseus cluster, made with an array including the orbiting radio telescope of the \textit{RadioAstron}\cite{Kardashev2013} mission. The obtained image transversely resolves the edge-brightened jet in 3C\,84 only 30\,microarcseconds from the core, which is ten times closer to the central engine than what has been possible in previous ground-based observations\cite{Nagai2014}, and it allows us to measure the jet collimation profile from $\sim 10^2$ to $\sim 10^4$ gravitational radii from the black hole. The previously found\cite{Nagai2014}, almost cylindrical jet profile on scales larger than a few thousand $r_\mathrm{g}$ is now seen to continue at least down to a few hundred $r_\mathrm{g}$ from the black hole and we find a broad jet with a transverse radius of $\gtrsim 250$\,$r_\mathrm{g}$ at only 350\,$r_\mathrm{g}$ from the core. If the bright outer jet layer is launched by the black hole ergosphere, it has to rapidly expand laterally on scales $\lesssim 10^2$\,$r_\mathrm{g}$. If this is not the case, then this jet sheath is likely launched from the accretion disk. 
\end{abstract}

\vspace{0.5cm}

3C\,84 (NGC\,1275, Per~A) is a Fanaroff-Riley type I radio galaxy at redshift of 0.0176. It has a two-sided parsec-scale jet and a strong nuclear emission resolved in a one-sided sub-parsec scale structure\cite{Walker1994,Nagai2014}. Due to its brightness and proximity 3C\,84 is an ideal laboratory to study the jet structure and origin. At the redshift of 3C\,84, the angular scale of one milliarcsecond (mas) corresponds to only 0.344\,pc. With the assumed mass of $\sim 2 \times 10^9$\,M$_{\odot}$ for its central supermassive black hole (SMBH), this means that 1\,mas $\approx 3.58 \times 10^3$\,$r_\mathrm{g}$, where $r_\mathrm{g}$ is the black hole's gravitational radius (see Methods). 

We observed 3C\,84 on September 21, 2013 with a very long baseline interferometry (VLBI) array consisting of a global network of ground radio telescopes and the 10-metre Space Radio Telescope (SRT) of the \textit{RadioAstron} space-VLBI mission\cite{Kardashev2013}. The highly elliptical orbit of the SRT with an apogee height of 360\,000\,km provides baselines that are up to thirty times longer than what is possible on the Earth, thus allowing a significant improvement in the angular resolution over ground-based arrays. For 3C\,84, we succesfully detected the interferometric signal between the SRT and the ground array up to a baseline length of about 8\,Earth diameters. At the observing frequency of 22.2\,GHz this corresponds to only 27\,microarcsecond ($\mu$as) fringe spacing on the sky. This is the smallest size scale our data are sensitive to by the Rayleigh criterion and it corresponds to about hundred gravitational radii in 3C\,84.  

Figure~1  presents our 22\,GHz space-VLBI image of the innermost one parsec of 3C\,84 at an angular resolution of $0.10 \times 0.05$\,mas (PA=0$^\circ$). The bright and compact emission at about 2.2\,mas north of the image reference center is identified with the radio core, from where a faint and short counter-jet and a brighter 3-mas-long jet depart, toward the north and south directions, respectively. The main jet ends in a bright spot with a surrounding diffuse emission identified with the emission feature called "C3" in the previous studies\cite{Nagai2014}. This feature emerged from the core around year 2003, moves at speed of $<0.5c$ and appears to be the end point of the newly re-started jet where the flow strongly interacts with the external medium\cite{Suzuki2012,Nagai2017}. An older, more slowly moving and diffuse emission feature, called "C2"\cite{Suzuki2012}, is visible on the western side of the jet. 

The limb-brightened jet connecting the core and C3 shows a large initial opening angle, followed by a rapid collimation to a quasi-cylindrical shape. Most AGN jets in VLBI images appear ridge-brightened, while limb-brightened jets are rare and have been reported only in a few nearby radio galaxies like Mrk\,501\cite{Giroletti2004}, M\,87\cite{Hada2016}, and Cygnus\,A\cite{Boccardi2016}, as well as in 3C\,84\cite{Nagai2014} itself. Figure~2 shows the inner core-jet region convolved with a 0.05\,mas circular beam and reveals that the jet remains strongly limb-brightened all the way to the core. Thanks to the 0.027\,mas fringe spacing roughly in the direction transverse to the jet, our space-VLBI observation resolves well the two limbs already at a distance $z_\mathrm{proj} \approx 0.03$\,mas from the core. This is a factor of ten closer to the central engine than what was possible with the earlier ground-based VLBI measurements at 43\,GHz (having East-West resolution of 0.13\,mas)\cite{Nagai2014} and corresponds to a de-projected distance of $z \approx 350$\,$r_\mathrm{g}$ assuming a jet inclination angle of 18$^\circ$\cite{Tavecchio2014} (see Methods). The presence of the limb-brightened structure so near to the jet origin confirms that it is an intrinsic property of the jet.

The large intensity ratio between the bright outer layer (hereafter "sheath") and the dim central part of the jet (hereafter "spine") can be explained either by a specific transverse velocity structure of the flow\cite{Komissarov1990} or by intrinsic emissivity differences between the spine and the sheath -- or both. Magnetohydrodynamic simulations of black hole ergosphere-driven jets show a slow outer jet layer in addition to a fast central part\cite{McKinney2006}. At intermediate-to-large jet inclination angles ($\theta \gtrsim 10^\circ$), it is possible to have a situation in which relativistic Doppler boosting effect is stronger for the low-velocity sheath than for the high-velocity spine, thus producing images where the sheath is apparently the brightest jet region. On the other hand, one also expects higher mass loading of the sheath due to its interaction with the external medium as is evident by the deceleration of the jets on the large scale. If the electrons at jet boundary layer can be accelerated by some mechanism, such as shear acceleration\cite{Stawarz2002}, the intrinsic emissivity in the sheath can exceed that of the spine. The brightening of the spine as the jet approaches C3 is visible in Fig.~1 and it can indicate either slowing down of the flow or increased emissivity caused by particle acceleration in C3, which likely contains strong shocks set up in the interface of the jet and the interstellar medium.

%

The bright and compact sheath sides provide a well-determined outline of the jet, thus allowing a very robust measurement of its collimation profile. The peak-to-peak width ($2r$) of the edge-brightened jet at different distances from the core is shown in Fig.~3. The jet appears to be surprisingly wide already at the nearest point to the radio core where we can measure the jet radius using the limb-brightened structure, $z_\mathrm{proj}=0.03$\,mas ($z \approx 350$\,$r_\mathrm{g}$ de-projected). Here $r=0.07$\,mas, corresponding to $r \approx 250$\,$r_\mathrm{g}$ or more depending on the black hole mass assumption. Assuming that the jet origin coincides with the location of the 22\,GHz core, the apparent jet opening angle $\alpha_\mathrm{o} = 2 \arctan (r/z_\mathrm{proj}) = 130^\circ \pm 10^\circ$. This is the largest opening angle ever observed in any astrophysical jet (for M\,87 $\alpha_\mathrm{o} \sim 100^\circ$ has been measured\cite{Hada2016}). The corresponding intrinsic opening angle is $\alpha_\mathrm{i} = 2 \arctan [ \tan (\alpha_\mathrm{o}/2) \sin \theta ] \approx 70^\circ$. Despite this large initial opening angle, the collimation profile between $z=350$\,$r_\mathrm{g}$ and $z=8\,000$\,$r_\mathrm{g}$ is almost cylindrical with $r \propto z^{0.17\pm0.01}$. This quasi-cylindrical profile has been seen in an earlier study on scales larger than a few thousand $r_\mathrm{g}$\cite{Nagai2014}, but now there is clear evidence that it exists already at a few hundred $r_\mathrm{g}$ from the central engine. As is also apparent in Fig.~2, this implies a strong collimation of the jet inside a few hundred $r_\mathrm{g}$ from the core. 

AGN jets are likely powered by magnetic fields extracting either the energy of the accreting matter\cite{Blandford1982} or the rotational energy of the spinning black hole itself\cite{Blandford1977}. Both theoretical arguments and recent computer simulations\cite{Tchekhovskoy2012} favour jet launching from the black hole ergosphere (so-called Blandford-Znajek (BZ) mechanism\cite{Blandford1977}) -- especially in those AGN that have geometrically thick, radiatively inefficient accretion flows (RIAFs). With the Eddington ratio of 0.4\%, 3C\,84 is considered as a RIAF, although just barely\cite{Plambeck2014}. If the jet streamlines are anchored at the event horizon, the maximum width of the jet close to the central engine is restricted. Our measured jet radius of $\gtrsim250$\,$r_\mathrm{g}$ at $z \approx 350$\,$r_\mathrm{g}$ from the central engine places significant constraints for this scenario. Figure~3 shows two theoretical streamlines from force-free, steady-state jet solutions anchored at the horizon radius on the equatorial plane as the outermost streamlines that can touch the horizon. It is clear that the limb-brightened jet structure is much wider than the genuine parabolic (Blandford--Znajek-type: $r \propto z^{0.5}$ at $r \gg r_\mathrm{g}$) streamline. The measured data points remain just below the quasi-conical ($r \propto z^{0.98}$ at $r \gg r_\mathrm{g}$) outermost streamline. This implies that, while the streamlines of the jet sheath may in principle connect to the horizon, this is only possible if there is a rapidly laterally expanding flow on scales $\lesssim 10^2$\,$r_\mathrm{g}$. 
Hence, while an ergosphere-launched jet\cite{Blandford1977} still remains possible in the case of 3C\,84, we should consider the possibility that the jet sheath is launched from the accretion disk\cite{Blandford1982}. We note that this does not exclude the existence of a magnetically collimated, ergosphere-launched core \emph{inside} the sheath.

The quasi-cylindrical jet structure in 3C\,84 is in sharp contrast with the two other jet collimation profiles that have been measured up to now. Both of these are nearly parabolic: M\,87 has  $r \propto z^{0.56\pm0.01}$ between $z= 200$\,$r_\mathrm{g}$ and $z = 4 \times 10^5$\,$r_\mathrm{g}$ (dashed line in Fig.~3; Nakamura et al. submitted to \apj) and Cygnus~A has $r \propto  z^{0.55\pm0.07}$ between $z = 500$\,$r_\mathrm{g}$ and $z = 10^4$\,$r_\mathrm{g}$\cite{Boccardi2016}. Since the outline of the relativistic MHD jets is determined by confinement due to external medium, this difference in collimation profiles implies differences in the environments, where the jets propagate. 

If the pressure in the external medium decreases as $p_\mathrm{ext} \propto z^{-b}$ with $b < 2$, there exists an equilibrium solution\cite{Lyubarsky2009} with a collimation profile $r \propto z^{b/4}$. The observed quasi-cylindrical collimation profile, $r \propto z^{0.17}$, could therefore be produced by a very shallow pressure profile of the external medium, $b \lesssim 1$. The corresponding density profile at $z \lesssim 1$\,pc should be then close to flat, i.e., $\rho_\mathrm{ext} \propto z^{-k}$ with $k = b-1 \approx 0$. Since both the spherical Bondi accretion model and the advection-dominated accretion flow model have $k=3/2$, these nearly free-falling accretion flows cannot explain the observed collimation profile\cite{Nakamura2013}. 
If it is a disk-like accretion flow that confines the jet, then its scale height should be at least $\sim 10^4 r_\mathrm{g} \sim 0.8$\,pc in order to explain our observations. However, having a flat density profile along the inner edge of a geometrically thick disk seems also unlikely. Hence, it is most likely that the jet is not in a pressure equilibrium with the accretion flow or other stratified components of the interstellar medium (ISM).

One obvious difference between the jets in M\,87 and 3C\,84, which can provide hints regarding the origin of the collimation profile in 3C\,84, is the restarted nature and young age of the latter\cite{Nagai2017}. The dynamical age of the feature C3, the head of the restarted jet, is only $\sim 10$\,yr at the time of our observation\cite{Suzuki2012}. Kiloparsec scale jets are known to create cavities with an almost uniform pressure environment, which can recollimate the flow into a cylindrical shape before it enters the leading hot spot\cite{Komissarov1998}. The feature C3 likely corresponds to a parsec-scale analogue of kiloparsec scale hot spots\cite{Nagai2017}, and the restarted jet may have thus recollimated already very close to the central engine. This is supported by \textit{RadioAstron} observations at 5\,GHz, which show evidence for low-intensity cocoon emission surrounding the 22\,GHz jet (Savolainen et al., in prep.). Hence, the jet in 3C\,84 is probably not shaped by the underlying stratified ISM, but by the shocked material of the cocoon, and the oscillations of the jet width beyond 8\,000\,$r_\mathrm{g}$ in Fig.~3 may be manifestations of this same interaction. Finally, we note that the dynamical age of the restarted jet is less than what is likely needed for the relaxation of the system (the sound-crossing time is $\gtrsim 10$\,yr for a sphere of 1\,pc radius even if one assumes a maximum sound speed of 0.3$c$ for the ambient medium) and we may not be seeing the final structure of the jet. Future observations of 3C\,84 may therefore give a unique record of the early evolution of a restarted jet in an active galaxy.

\begin{methods}

\subsection{Observations and data reduction}
We give here a concise description of the space-VLBI observations and applied data reduction procedures. A more detailed account is given in a companion paper, which also discusses the observations made at other frequencies during the same observing run (Savolainen et al. in prep.).

3C\,84 was observed by the \textit{RadioAstron}\cite{Kardashev2013} Space Radio Telescope (SRT) and an array of ground radio telescopes in a VLBI mode around the perigee passage of the SRT from 2013-09-21 15:00 UT to 2013-09-22 13:00 UT (observation codes raks03a for \textit{RadioAstron} and GS032A for the ground array). Projected space baselines from 0.2 to 10.4 Earth diameters in length were sampled during the observation. The SRT recorded simultaneously left circularly polarized (LCP) signals from both the C-band (4.836\,GHz) and K-band (22.236\,GHz) receivers. The total recorded bandwidth was 32\,MHz at each band. The SRT recorded 44 ten-minute long blocks of data, hereafter called "scans". There were 70$-$90\,min long gaps between every 3--4 scans in order to allow the satellite's high gain antenna motor to cool down. The global ground array observed the source continuously, when it was above the horizon. The \textit{RadioAstron} data were recorded by tracking stations in Puschino, Russia and in Green Bank, USA. 

The ground array consisted of 29 radio telescopes, of which 24 produced data that were successfully correlated. The array was split in two parts during the \textit{RadioAstron} observing scans: five Very Long Baseline Array (VLBA) antennas (Brewster, Kitt Peak, North Liberty, Pie Town, St.~Croix), Green Bank Telescope (GBT), Shanghai, Kalyazin, Onsala, Noto, Jodrell Bank, Westerbork and Hartebeesthoek observed only at 4.8\,GHz, while Korean VLBI Network (KVN) antennas Yonsei and Ulsan, as well as five VLBA antennas (Fort Davis, Hancock, Los Alamos, Mauna Kea, Owens Valley), the phased Karl~G.~Jansky Very Large Array (VLA), Medicina and Yebes observed at 22.2\,GHz. Effelsberg switched between the bands, observing 50\% of the time at each. Frequency-agile VLBA antennas and the phased VLA observed additionally at 15.4, 22.2 and 43.2\,GHz during the \textit{RadioAstron} cooling gaps. The recorded baseband data were correlated at the Max-Planck-Institut f\"ur Radioastronomie using the DiFX correlator modified for space-VLBI application by implementing a rigorous model for the path delay of an interferometer with an orbiting element according to the general relativity\cite{Bruni2014}. 

Finding interferometric signal (i.e. "fringes") on the baselines to the orbiting antenna is challenging due to the typically low signal-to-noise (SNR) ratio of the fringes and the large parameter space that needs to be searched due to uncertainties in the \textit{a priori} orbit reconstruction of the SRT. The post-correlation fringe search was carried out in two parts. First, a coarse search was performed with the PIMA software\cite{Petrov2011} (\url{http://astrogeo.org/pima/}) which processes baselines individually. We selected solutions that had false detection probability less than 0.1\%, and determined the large residual group delay, fringe rate, and fringe acceleration that are due to the uncertainties in the \textit{a priori} orbit. The PIMA search resulted in fringe detections up to baseline lengths of 6.9\,$D_\mathrm{Earth}$ at 4.8\,GHz and up to 2.8\,$D_\mathrm{Earth}$ at 22.2\,GHz. In the second step of fringe-fitting we refined the model derived in the previous step using the Astronomical Image Processing System (AIPS; \url{http://www.aips.nrao.edu}) task \textsc{fring}, which allows combining data from multiple ground telescopes in a global solution. Such a combined solution is equivalent to phasing up the ground array antennas and it increases the sensitivity with respect to the baseline-based fringe search in the step one\cite{Kogan1996}. 

3C\,84 has a complex, extended structure in VLBI scales, which causes additional noise in the global fringe fitting solutions, if it is not taken into account. In order to remove the effect of the source structure, we first imaged the source by using only the ground baselines. The resulting image (see Supplemental Information) was then used as an input model for global fringe-fitting of the full data set, including the space baselines, in AIPS \textsc{fring}. Fringe-fitting of the SRT data in AIPS was performed in an iterative manner: on the first round a moderately large search window with a detection threshold of SNR=5 was used, and on the second round the window was narrowed down to $\pm100$\,ns in delay and $\pm25$\,mHz (4.8\,GHz) or $\pm50$\,mHz (22.2\,GHz) in rate around the values interpolated from the neighbouring solutions (combining detections at both bands) and the SNR threshold was lowered to 3.1 that corresponded to false detection rates below 0.1\% (4.8\,GHz) and 0.2\% (22.2\,GHz). We note that the ground array data were fringe-fitted at 2\,min solution interval before carrying out the fringe search on the space baselines at 10\,min solution interval. Therefore, much of the atmospheric phase fluctuations were removed before SRT fringes were searched, thus allowing longer integration times.

The AIPS fringe search yielded space-baseline fringe detections for 33 scans at 4.8\,GHz with the longest baselines being 7.8\,$D_\mathrm{Earth}$ to Effelsberg and 8.1\,$D_\mathrm{Earth}$ to GBT. At 22.2\,GHz space baseline fringes were detected for 12 scans with the longest baselines being 7.6\,$D_\mathrm{Earth}$ to Effelsberg and 8.1\,$D_\mathrm{Earth}$ to VLA. The measured visibilities at 22\,GHz cover a range in ($u,v$) radius from about 4\,M$\lambda$ to 7.7\,G$\lambda$. The visibilities on the space baselines comprise 5.6\% of the total number of K-band visibilities after the fringe search stage.

The gain amplitude calibration was performed in a standard manner using measured system temperatures and gain curves\cite{Kovalev2014}. Editing, imaging and self-calibration of the data were carried out in Difmap. While the ground array antennas were self-calibrated down to 10\,s averaging interval in phase and to 30\,s averaging interval in amplitude, the SRT was self-calibrated using longer solution intervals: 2\,min in phase and the whole observing length in amplitude. This is important in order to prevent spurious flux from being generated from the noise on the longest space-baselines that have weaker constraints from closure phases. The corrections in the amplitude for the SRT were modest, 15$-$20\,\%.

Since the aim of the space-VLBI observations is to obtain the highest possible angular resolution, we give more weight to the space baselines in imaging than what the usual natural weighting scheme does\cite{Murphy2000}. In Difmap, we selected uniform weighting with a bin size of 5 pixels in the $(u,v)$ grid combined with an additional weighting by the visibility errors to the power of $-1$ as this gave a good balance between the angular resolution and noise in the final image. We note that the data were also independently imaged in AIPS and the results agree well.

The full resolution image at 22.2\,GHz has a beam size of $300 \times 50$\,$\mu$as (PA=22$^\circ$) when using above-described weighting. The highly elongated beam is due to the $(u,v)$ coverage of space baselines being in a narrow range of position angles around $\sim 100^\circ$. In order to make the image easier to interpret by eye, Fig.~1 uses a more symmetrical restoring beam of $100 \times 50$\,$\mu$as, i.e., the image is super-resolved by a factor of three in one direction. This amount of super-resolution was found in a recent study to give minimum errors in the CLEAN image reconstruction of simulated data sets\cite{Akiyama2017}. Also, a comparison of source structures between images made with different $(u,v)$ weighting functions and different restoring beams shows a good agreement.


\subsection{Measurement of the jet collimation profile}
To measure the jet width as a function of distance from the core, we used the image which was convolved with a 0.05\,mas circular beam and had a pixel-size of 0.002\,mas. All the distances were measured from the image peak flux density position (assumed to be the core position at 22\,GHz) and the center of the well-resolved jet (see next subsection for a discussion on a possible core-shift). To measure the jet width we obtained in AIPS multiple slices perpendicular to the jet direction. The first slice is at a distance of 0.03\,mas, where the jet is already well-resolved in transverse direction. 

The eastern side of the jet sheath is marginally resolved within the first 0.5\,mas from the core with a deconvolved full-width-at-half-maximum (FWHM) $< 15$\,$\mu$as, while the western side is slightly more extended with a FWHM of $22 \pm 8$\,$\mu$as. The width of both sides constantly increases as a function of distance from the core, reaching $40 \pm 8$\,$\mu$as at $\sim 1$\,mas from the core before merging with C3. These bright and narrow jet limbs provide a well-determined outline of the jet, thus allowing us to accurately measure the jet width (2$r$) as the peak-to-peak distance between two Gaussians fit to the bright East and West edges of the jet. The two Gaussian profiles are always well separated and the brightness in the central region of the jet is low, allowing good fits. Uncertainties in the jet width reported in Fig.~3 are in the range of 0.01$-$0.02\,mas (1\,$\sigma$) and have been estimated from the reported uncertainties in the Gaussian fit, and comparing results with a small shift (2$-$4\,pixels) of the slice position. Uncertainties in the distance from the core are very small -- of the order of 2\,pixels. Therefore the uncertainty in the observed opening angle is relatively small, and we estimate $\alpha_o = 130^\circ \pm 10^\circ$.

\subsection{Possible core-shift}
The bright, upstream-most emission feature in AGN jets is known as the "core" and it is usually identified as the location where the optical depth due to synchrotron self-absorption is $\sim 1$. This location is frequency-dependent, and the resulting measurable phenomenon is known as "core-shift"\cite{Lobanov1998}. In our analysis, we have assumed that the 22\,GHz core is coincident with the jet origin. However, if there is significant core-shift, all the jet width measurements would move to the right in Fig.~3 by the corresponding amount.

We can constrain the possible core-shift in 3C\,84 at 22\,GHz thanks to detection of the counter-jet in our image, which is also in agreement with the recent lower resolution images\cite{Fujita2017}. Directly measuring the gap between the jet and the counter-jet in the 22\,GHz image is not obvious, because one would need to subtract the strong central component and the residuals suffer from dynamic range problems. However, moving to the East of the map peak in Fig.~2, we see clearly that the radio structure is forking and there is emission to the north and to the south with respect to the peak position. We interpret this structure as the region where the jet and the counter-jet start.  Using \textsc{tvslice} in AIPS, and looking at \textsc{clean} components in the high resolution image, we can estimate that the gap in between the two jet regions is $0.05 \pm 0.02$ mas. Since the central engine should be located in between the jet and the counter-jet we conclude that the core-shift at 22\,GHz should not be more than 0.03\,mas in north-south direction. We tried the same procedure with the western side of the jet, but there the uncertainties are too large, since the jet is significantly brighter than the counter-jet.

Assuming a core-shift of 0.03\,mas, the jet opening angle becomes smaller. The first point where the jet width is accurately measured from the edge-brightened structure would now be located at $z_\mathrm{proj} = 0.06$\,mas from the jet origin (instead of $z_\mathrm{proj} = 0.03$\,mas) and the corresponding apparent opening angle would be $\sim 100^\circ$ and the intrinsic opening angle would be $\sim 40^\circ$. This is still a large value and the jet collimation profile and the corresponding discussion in the text are only marginally affected by the possible core-shift. In Fig.~3 we show the effect of possible 0.03\,mas core-shift by right-side horizontal error bars.

\subsection{Jet orientation and black hole mass estimates}

The redshift of 0.0176 corresponds to the angular scale of 0.344\,pc/mas for 3C\,84 assuming a $\Lambda$CDM cosmology with H$_0 = 70.7$\,km\,s$^{-1}$ Mpc$^{-1}$, $\Omega_\mathrm{M} = 0.27$ and $\Omega_\mathrm{\Lambda} = 0.73$. For converting the measured angular distances along the jet to de-projected linear distances in units of gravitational radii, we need to adopt estimates for the jet inclination angle and black hole mass. 

The jet orientation with respect to the line-of-sight in 3C\,84 has been extensively discussed in the literature, but it is far from certain. A relatively large orientation angle of $30^\circ$$-$$55^\circ$ has been estimated by comparing the brightness and distance ratios between the southern (main jet) and the northern jet (counter-jet) on 10$-$20\,mas scale\cite{Walker1994,Asada2006}. On the other hand, the proper motion ratio of the two jets indicates a much smaller viewing angle of 11$^\circ$\cite{Lister2009b}. This discrepancy is not so surprising, considering that the counter-jet emission on this scale is strongly affected by free-free absorption\cite{Walker1994} and that the jet velocities are slowed down by the interaction with the dense and likely clumpy medium. For these reasons the intrinsic symmetry in size, speed and brightness can be strongly affected by external, non-symmetric conditions, which renders these inclination estimates uncertain. Recently, a detection of a diffuse emission region 2\,mas north of the core was reported and an inclination of 65$^\circ$ was derived assuming that this feature is a counterpart to C3\cite{Fujita2017}. However, since no compact emission similar to the hot spot in C3 was detected for this northern component, it is unclear whether it indeed corresponds to the end point of the restarted jet. Again, the same caveats regarding the symmetry of the jet in a dense and clumby medium apply also to this estimate.
%
We note that we cannot use the brightness ratio between the jet and the counter-jet in our Fig.~1 to constrain the jet inclination for 3C\,84, since the jet is a young, strongly evolving structure with a large variability in its brightness, morphology and proper motion\cite{Suzuki2012,Nagai2017}. Since the structures visible on the jet and the counter-jet side have different ages, no comparison is possible.

On the other hand, the broadband spectral energy distribution (SED) of 3C\,84 suggests smaller viewing angles for its jet\cite{Aleksic2014,Tavecchio2014}. Specifically, the SED can be satisfactorily reproduced in the framework of the spine-layer model\cite{Tavecchio2014}, if the viewing angle is smaller than 20$^\circ$. Considering that there indeed is a strongly stratified jet structure visible in Fig.~1, an inclination of 18$^\circ$ was adopted for the discussion in this Letter. 

If we assume that the limb-brightened structure can be explained by the inner spine of the jet being faster than the outer sheath\cite{Komissarov1990}, we can use the brightness ratio between the sheath and the spine to place some additional constraints on the jet orientation and velocity. The observed brightness ratio between the sheath and the spine in 3C\,84 is about 20 within the innermost 0.55\,mas from the core. For a jet inclination of $18^\circ$, the maximum Doppler boosting factor $\delta = 3.23$ is obtained with a flow velocity of 0.943\,$c$ (corresponding to a Lorentz factor of $\Gamma = 3$), which is still a reasonable value for a jet sheath\cite{Tavecchio2014}. Assuming this velocity for the sheath and furthermore assuming that both the sheath and the spine have the same intrinsic emissivity, we find that a spine velocity $\Gamma \gtrsim 20$ is needed to produce the observed brightness ratio. If the inclination angle is significantly larger than 18$^\circ$, the spine velocity can be lower, but still $\Gamma \gtrsim 10$ is required. If a transverse velocity structure is the only factor determining the limb-brightened appearance of the jet, the spine must have thus accelerated to a velocity of $\Gamma \gtrsim 10$ already within the first few hundred $r_\mathrm{g}$. The observed flux ratio also excludes inclination smaller than about $10^\circ$. If the spine has a lower relativistic particle density than the sheath, these constraints can be relaxed. 

While we adopt an intermediate inclination angle of 18$^\circ$, it is not possible to exclude a large inclination of $\sim 45^\circ$, either. Such a viewing angle would make the measured intrinsic opening angle larger by a factor of $\sim2$ and it would also change the assumed black hole mass as explained below. In Fig.~3 we therefore present the jet collimation profile assuming both a moderate jet inclination of 18$^\circ$ and a large inclination of 45$^\circ$. 

The mass of the supermassive black hole (SMBH) at the center of the NGC\,1275, the optical counterpart of 3C\,84, has been discussed in the literature based on the molecular gas dynamics in the centre of the galaxy\cite{Wilman2005,Scharwaechter2013}. The reported values are dependent on the inclination of the molecular disc rotation axis, which has been assumed to match that of the radio jet axis.  Recent high-resolution near-IR integral field spectroscopy\cite{Scharwaechter2013} indicates a central mass of $8 \times 10^8$\,M$_{\odot}$ for a disk inclination of $45^\circ\pm10^\circ$ with a global lower limit of $5 \times 10^8$\,M$_{\odot}$. A smaller inclination angle, like in our discussion, implies a significantly higher central mass from the same observations (more than $2 \times 10^9$\,M$_{\odot}$). However, since it is possible that the jet inclination can differ from the inclination of the molecular disk, and also since the direct observations do not suggest a highly inclined disk\cite{Scharwaechter2013}, we consider $2 \times 10^9$\,M$_{\odot}$ as the upper limit of the SMBH mass for our jet inclination of 18$^\circ$. This yields 1\,$r_\mathrm{g} = 0.96 \times 10^{-4}$\,pc $= 2.793 \times 10^{-4}$\,mas or 1\,mas $= 3.58 \times 10^3$\,$r_\mathrm{g}$, where $r_\mathrm{g}$ is the gravitational radius. We have assumed this scaling for our discussion in order to be consistent with the assumed jet inclination of 18$^\circ$.

As mentioned earlier, we cannot exclude the possibility of a large jet inclination of 45$^\circ$. In this case the SMBH mass would of course be correspondingly lower, $8 \times 10^8$\,M$_{\odot}$, with a lower limit of $5 \times 10^8$\,M$_{\odot}$\cite{Scharwaechter2013}. Interestingly, while the lower black hole mass scales up the jet width in gravitational radii by a factor of 2.5 (or by a factor of four if the lower limit for the mass is assumed), the de-projected distances along jet change much less - by only 10\% (or by a factor of 1.7, if the lower limit for the mass is assumed). This is because the change in the foreshortening factor, $\sin 45^\circ / \sin 18^\circ = 2.3$, counters the change in the BH mass scaling. This can be seen in Fig.~3, which shows the jet collimation profile also for the cases of $i=45^\circ$ and $\mathrm{M_{BH}} = 8 \times 10^8$\,M$_{\odot}$ or $5 \times 10^8$\,M$_{\odot}$.
 
\end{methods}


\vspace{0.8cm}
\bibliographystyle{naturemag}
\bibliography{radioastron}


\begin{addendum}
 \item We thank E.~Ros and the anonymous referees for useful comments. The RadioAstron project is led by the Astro Space Center of the Lebedev Physical Institute of the Russian Academy of Sciences and the Lavochkin Scientific and Production Association under a contract with the State Space Corporation ROSCOSMOS, in collaboration with partner organizations in Russia and other countries. The National Radio Astronomy Observatory is a facility of the National Science Foundation operated under cooperative agreement by Associated Universities, Inc. The European VLBI Network is a joint facility of independent European, African, Asian, and North American radio astronomy institutes.
 The KVN is a facility operated by the Korea Astronomy and Space Science Institute. The KVN operations are supported by KREONET (Korea Research Environment Open NETwork) which is managed and operated by KISTI (Korea Institute of Science and Technology Information). This work was partially supported by the National Research Council of Science \& Technology (NST) granted by the International joint research project (EU-16-001).
 This research is based on observations correlated at the Bonn Correlator, jointly operated by the Max Planck Institute for Radio Astronomy (MPIfR), and the Federal Agency for Cartography and Geodesy (BKG).
 T.S.\ was funded by the Academy of Finland projects 274477 and 284495. Y.Y.K., M.M.L., K.V.S., P.A.V.\ were supported by the Russian Science Foundation (project 16-12-10481). S.S.L.\ was supported by the National Research Foundation of Korea (NRF) grant funded by the Korean government (MSIP; No.~987 NRF-2016R1C1B2006697).
 \item[Author contributions] G.G., T.S. and M.O. coordinated the research, carried out the image analysis, and wrote the manuscript. T.S., Y.Y.K., K.V.S., S.-S.L., B.W.S., J.A.Z., planned and organized the space-VLBI imaging experiment including the ground array. G.B. correlated the VLBI data, with help from P.V., using the software tools developed by J.M.A. and L.P. Correlated VLBI data were calibrated by T.S. with contributions from M.M.L., while G.G., T.S., M.O. and Y.Y.K. imaged the data. The modelling was carried out by M.N., H.N., M.K. and M.G. All the authors contributed to the discussion of the data, its interpretation and provided comments on the manuscript. T.S. is the Principal Investigator of the \textit{RadioAstron} Nearby AGN Key Science Program.
 \item[Competing Interests] The authors declare that they have no competing financial interests.
 \item[Correspondence] Correspondence and requests for materials should be addressed to G.G.~(email: ggiovann@ira.inaf.it) and T.S.~(email: tuomas.k.savolainen@aalto.fi).
\end{addendum}


\clearpage

\begin{figure}[p]
\centering
\includegraphics[width=0.9\textwidth,trim=0cm 0cm 0cm 3cm]{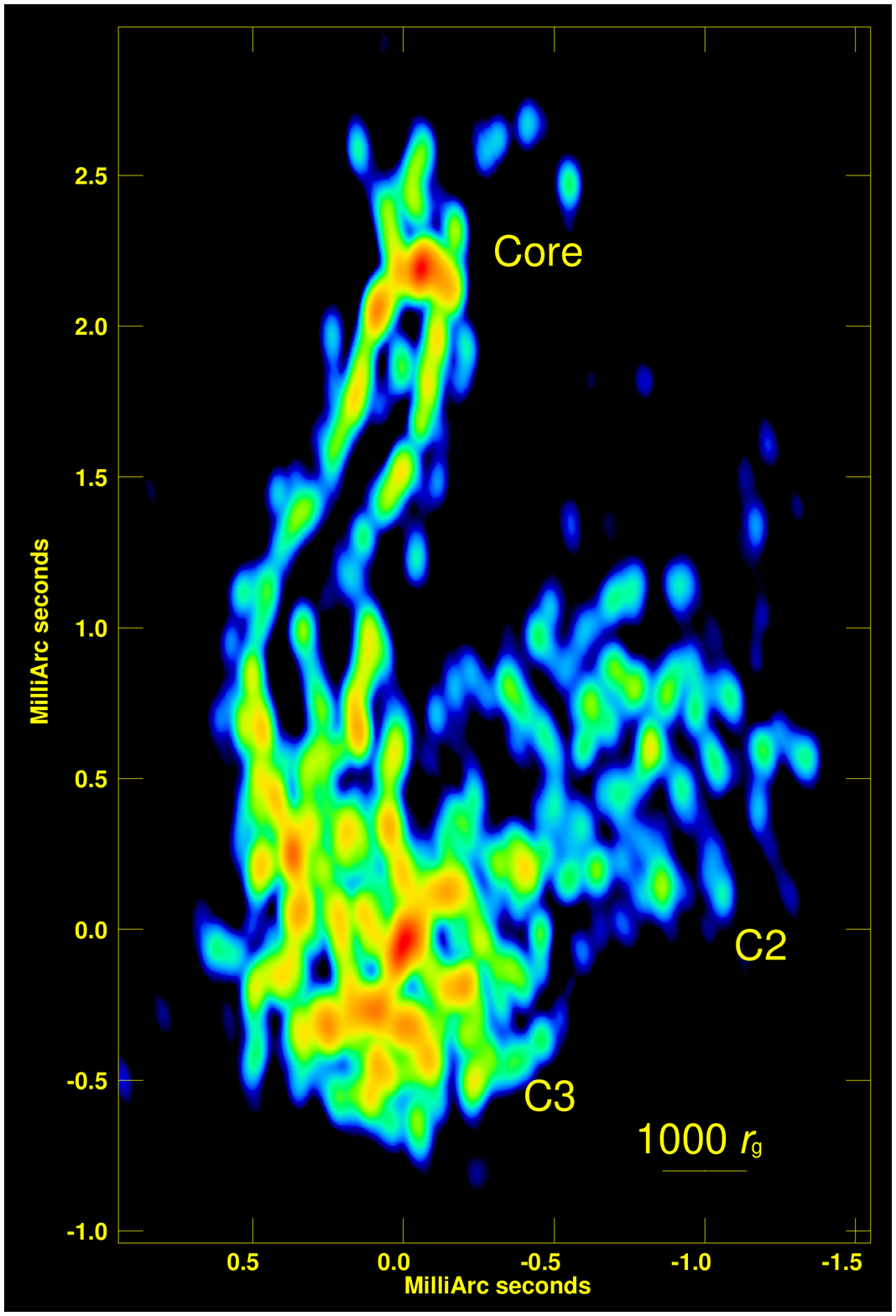}
\caption{Radio image of the central parsec in 3C\,84 obtained with the space-VLBI array. The half-power beam width (HPBW) is $0.10 \times 0.05$\,mas at PA=0$^\circ$. The noise level is 1.4\,mJy/beam and the peak intensity is 0.75\,Jy/beam. The radio core and emission features C2 and C3 (see text) are indicated in the image.}
\end{figure}

\begin{figure}[p]
\centering
\includegraphics[width=\textwidth,trim=0cm 4cm 0cm 4cm]{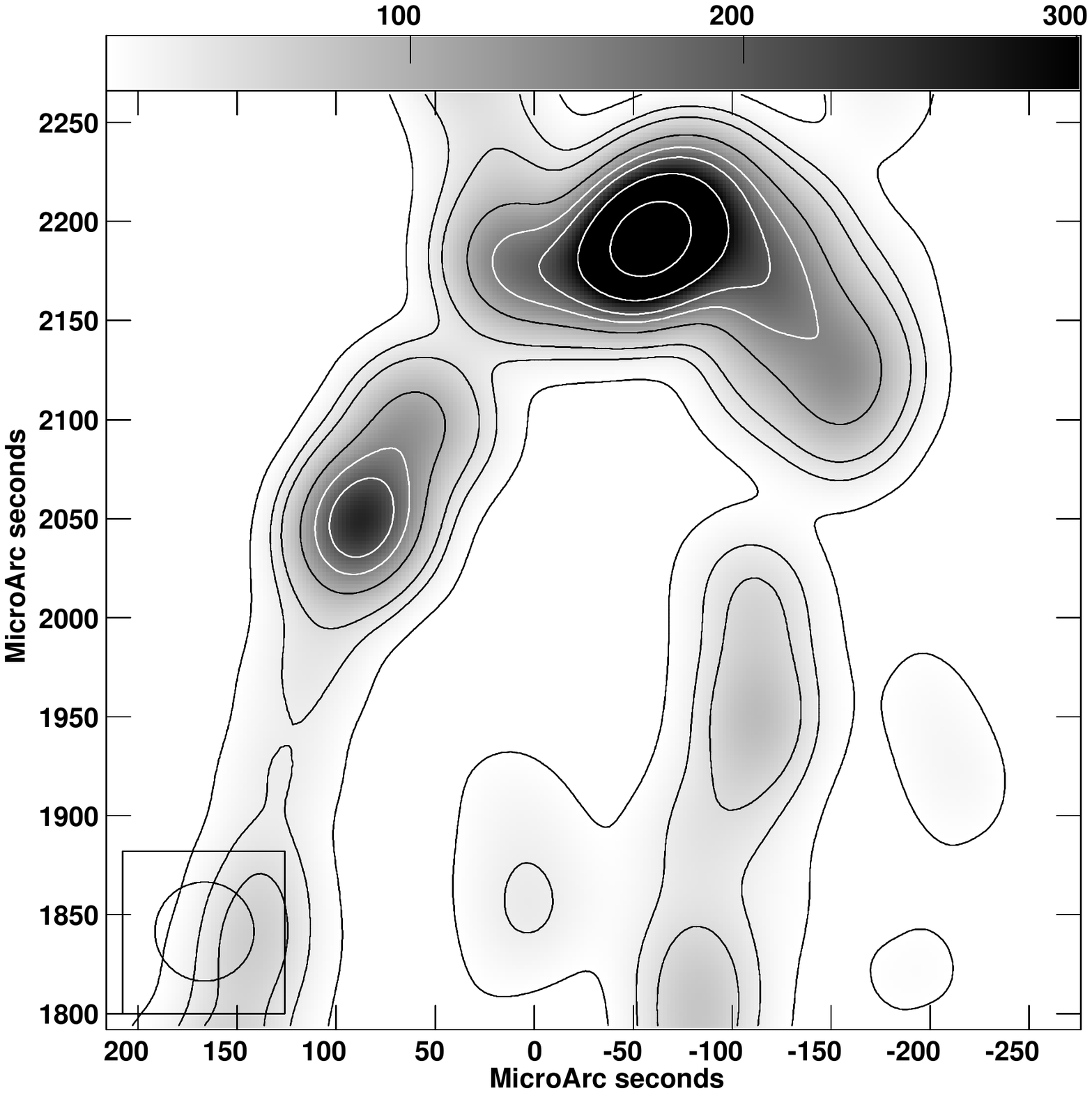}
\caption{Inner jet-core region at high angular resolution. The half-power beam width is $0.05 \times 0.05$\,mas. The noise level is 1.5\,mJy/beam and the peak intensity is 0.66\,Jy/beam. Contours are at 10, 30, 50, 100, 150, 200, 300, 500 mJy/beam.}
\end{figure}

\begin{figure}[p]
\centering
\includegraphics[width=0.75\textwidth,angle=-90]{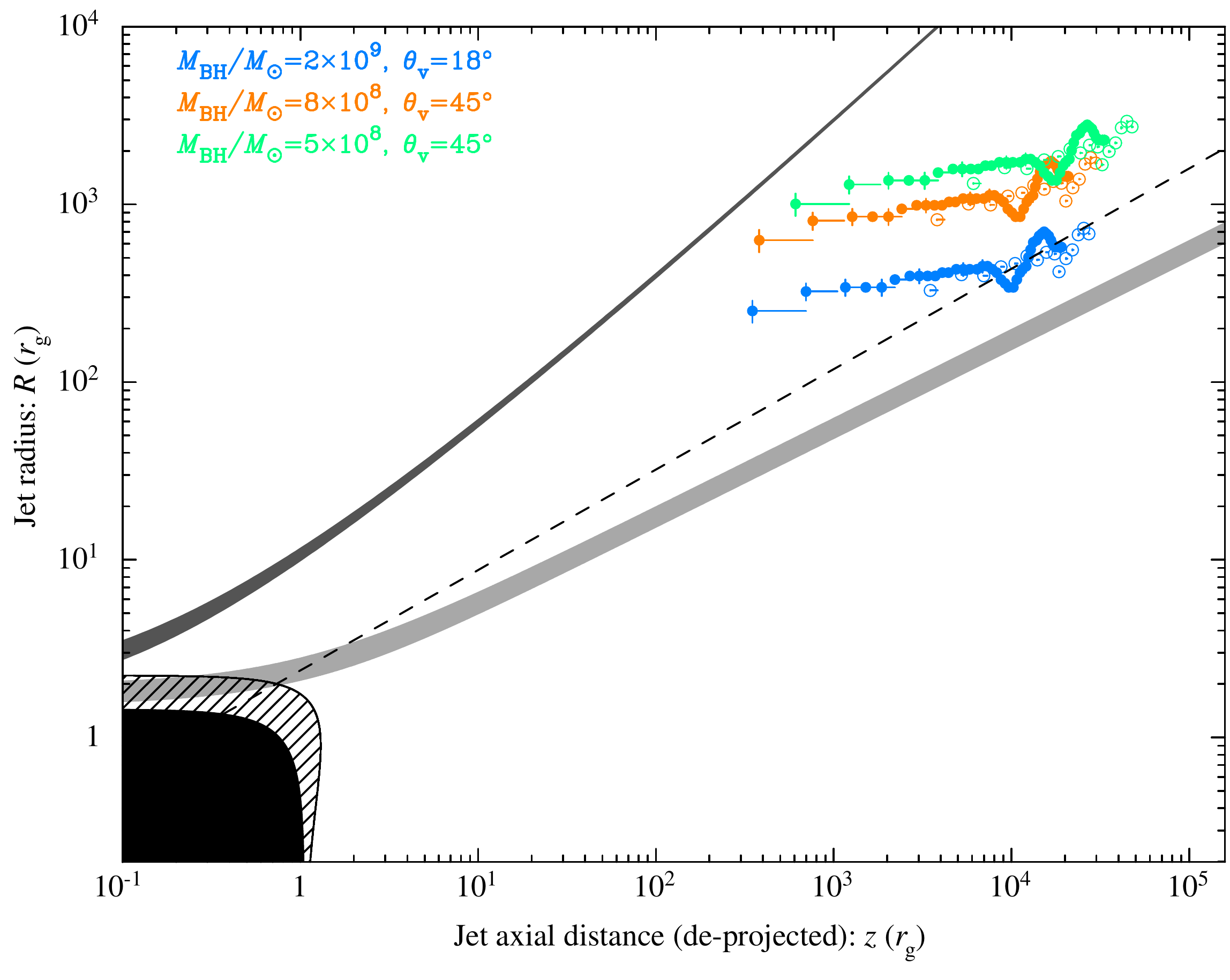}
\caption{Jet width as a function of de-projected distance from the central engine in units of gravitational radii. Filled points are 22\,GHz \textit{RadioAstron} data, empty circles are previously published VLBA data at 43\,GHz\cite{Nagai2014}. The data are plotted for three different assumptions of jet inclination angle and black hole mass (see Methods). Right-side horizontal error bars correspond to the uncertainty about a possible core-shift. The dashed line is the power-law fit to the collimation profile of M\,87 (Nakamura et al., submitted). The filled black region on the lower left corner denotes the black hole event horizon, while the hatched area represents the ergosphere for the black hole spin parameter $a=0.998$. The light gray area denotes the genuine parabolic streamline ($r \propto z^{0.5}$ at $r \gg r_\mathrm{g}$) of the force-free steady-state jet solution\cite{Blandford1977}, while the dark gray area denotes a quasi-conical outermost streamline ($r \propto z^{0.98}$ at $r \gg r_\mathrm{g}$) of the force-free steady-state jet solution\cite{Narayan2007}. In both streamlines a variation from $a=0.1$ (upper boundary) to $a = 0.998$ (lower boundary) is considered. Note that all the streamlines are anchored at the event horizon, $r_\mathrm{H} = r_\mathrm{g}(1 + \sqrt{1-a^2})$, with the maximum angle of $\theta = \pi/2$ in polar coordinates.}
\end{figure}



\end{document}